\documentclass[twocolumn]{article}
\pdfoutput=1

\usepackage{upgreek}  % not a standard package, may need to remove it for publication

\usepackage{hyperref}
\hypersetup{
    colorlinks=true,
    linkcolor=blue,
    citecolor=blue,     
    urlcolor=blue,
    }
    
\usepackage{geometry}

\usepackage{amsmath}
\usepackage{graphicx}

\usepackage{authblk}

\author[1,2,*]{Jonathan Peltier}
\author[3,*]{Weiwei Zhang}
\author[2]{Leopold Virot}
\author[1]{Christian Lafforgue}
\author[1]{Lucas Deniel}
\author[1]{Delphine Marris-Morini}
\author[1]{Guy Aubin}
\author[1]{Farah Amar}
\author[3]{Denh Tran}
\author[3]{Xingzhao Yan}
\author[3]{Callum G. Littlejohns}
\author[1]{Carlos Alonso-Ramos}
\author[3]{David J. Thomson}
\author[3]{Graham Reed}
\author[1]{Laurent Vivien}

\affil[1]{University Paris-Saclay, CNRS, Centre for Nanoscience and Nanotechnology (C2N), Palaiseau, 91120, France}
\affil[2]{University Grenoble Alpes, CEA, LETI, Grenoble, 38000, France}
\affil[3]{Optoelectronics Research Centre, Zepler Institute for Photonics and Nanoelectronics, Faculty of Engineering and Physical Sciences, University of Southampton, Southampton SO17 1BJ, UK}

\affil[*]{Corresponding author: jonathan.peltier@c2n.upsaclay.fr and Weiwei.Zhang@soton.ac.uk}

\title{High speed silicon photonic electro-optic Kerr modulation}

\begin{document}

\twocolumn[
  \begin{@twocolumnfalse}
  
\maketitle

\begin{abstract}
Electro-optic silicon-based modulators contribute to ease the integration of high-speed and low-power consumption circuits for classical optical communications or quantum computers.
However, the inversion symmetry in the silicon crystal structure inhibits the use of Pockels effect.
An electric field-induced optical modulation equivalent to a Pockels effect can nevertheless be achieved in silicon by the use of DC Kerr effect.
Although some theoretical and experimental studies have shown its existence in silicon, the DC Kerr effect in optical modulation have led to a negligible contribution so far.
This paper reports demonstration of high-speed optical modulation based on the electric field-induced linear electro-optic effect in silicon PIN junction waveguides.
The relative contributions of both plasma dispersion and Kerr effects are quantified and we show that the Kerr induced modulation is dominant when a high external DC electric field is applied.
Finally, the high-speed modulation response is analyzed and eye diagram up to 100 Gbits/s in NRZ format are obtained.
This work demonstrates high speed modulation based on Kerr effect in silicon, and its potential for low loss, quasi-pure phase modulation.
\end{abstract}

  \end{@twocolumnfalse}
  ]

\section{Introduction}
Integrated electro-optic modulators are key component in systems such as classical and quantum optical communications, photonics-based quantum computing and sensing.
These systems target high-speed and low power consumption optical modulators.
Silicon (Si) modulators, which rely primarily on the plasma dispersion \mbox{effect \cite{rahim_taking_2021}}, are intrinsically limited in speed due to their high RC \mbox{constant \cite{sinatkas_electro-optic_2021}.}
Si modulators relying on the Pockels effect could overcome these limitations to produce a fast and pure phase modulation.
Since silicon does not have a natural $\chi^{(2)}$ due to its centrosymmetric structure, such modulation cannot be achieved directly except by straining the crystal \mbox{lattice \cite{berciano_fast_2018}} leading to a low resulting Pockels coefficient.
The integration of \mbox{high-$\chi^{(2)}$} materials on the Si platform has been widely considered.
These include doped polymers, Barium Titanate (BTO) \cite{he_high_performance_2019}, Lead Zirconate Titanate (PZT) \cite{he_high_performance_2019} or lithium niobate \mbox{(LN) \cite{he_high_performance_2019}}.
These approaches require the development of hybrid or heterogeneous integration processes which increase the technology complexity.
An electro-optic modulation in Si can also be achieve through DC Kerr effect that electrically induces an \mbox{effective $\chi^{(2)}$} which can be hence exploited to vary the refractive index by applying an electrical modulation superimposed to a static field.
DC Kerr effect has been studied in bulk silica \cite{liu_measurement_2001}, bulk silicon \cite{chen_pockels_2008, zhu_investigation_2012}, silicon interface \cite{bodrov_terahertz-field-induced_2022}, bulk antiferromagnetic NiO \cite{chefonov_study_2022} and
in integrated platforms including \mbox{silicon-organic hybrid \cite{steglich_electric_2020}} silicon-rich nitride \cite{friedman_demonstration_2021}, silicon rich carbide \cite{chang_demonstration_2022} and in silicon \mbox{nitride \cite{zabelich_linear_2022}.}
It has also been studied in the silicon platform for electric field-induced (EFI) second-harmonic generation (EFISHG) \cite{timurdogan_electric_2017}, electro-optic (EO) modulation (EOM) \cite{bottenfield_silicon_2019, Jain_2019}, slow light regime \cite{xia_high_2022} and in cryogenic experiments \cite{chakraborty_cryogenic_2020}.
However, the high-speed EOM \mbox{in \cite{bottenfield_silicon_2019, Jain_2019, xia_high_2022}} using PN junctions led to a plasma dispersion effect that has a higher contribution to the modulation than the DC Kerr effect.
While the DC Kerr effect has been well studied in the DC regime, no assessment discriminating the contribution of the DC Kerr and plasma dispersion modulation in the dynamic regime has been reported to our knowledge.
This paper presents a comprehensive analysis of the DC Kerr effect induced in a PIN diode inserted in a silicon Mach-Zehnder Interferometer (MZI) in both static and dynamic regimes. Data transmission has been analyzed up to 100 Gbits/s in Non-Return-to-Zero (NRZ) format.
An experimental method has been developed to assess the relative contribution of plasma dispersion from the Kerr effect in the dynamic regime. 

The DC Kerr effect, also known as electric field-induced Pockels effect, originates from the third-order nonlinear susceptibility tensor \textbf{$\chi^{(3)}$} in presence of a static electric field.
The refractive index change induced by Kerr effect when a static electric field $F_{DC}$ and an RF field $F_{RF} \cos{\Omega t}$ are applied to the PIN junction is given by \cite{steglich_electric_2020}: 
\begin{equation}
    \begin{split}
        \Delta & n(t) = \\ & \frac{3\chi^{(3)}}{2n_{si}} ( F_{DC}^2 + \frac{1}{2} F_{RF}^2 + 2F_{DC}F_{RF} \cos{\Omega t} + \frac{1}{2}F_{RF}^2 \cos{2\Omega t} )
    \end{split}
\label{eq:delta_n_dc_kerr}
\end{equation}
with \mbox{$\Omega = 2\pi f$}, $f$ the RF frequency, \mbox{$n_{si}=3.48$} the silicon refractive index and \mbox{$\chi^{(3)}=2.8 \times 10^{-19}$ m$^2$.V$^{-2}$} at \mbox{$\lambda=1.55 \: \upmu$m}, for a silicon waveguide with a cross-section oriented along the crystallographic \mbox{axis [110] \cite{han_third-order-coef_2011,zhang-anisotropic-nonlinear-2007}}.
\mbox{Eq$.$ (\ref{eq:delta_n_dc_kerr})} exhibits three kinds of dependencies.
The first one corresponds to the static refractive index growing with the square of the field amplitudes that will be called later DC Kerr effect concerning $F_{DC}$.
The second one relies on an index modulation at an angular \mbox{frequency $\Omega$} which has its amplitude growing with the product of the DC and RF fields amplitudes. It will be called later electric field-induced (EFI) linear EO effect.
At last an index modulation at \mbox{a 2$\Omega$} component exhibits an amplitude growing with the square of the RF field amplitude alone. It will be called later quadratic EO effect.

\section{Results and discussions}

Static and dynamic studies are conducted to distinguish Kerr effects from that of plasma dispersion on the index variation in three different unbalanced Mach-Zehnder modulators (MZMs).
They consist of either PN or PIN junctions named PN, PIN2, PIN3 and their respective intrinsic region width are w=0, w=0.33 and 1.05 $\upmu$m (\mbox{Fig$.$ \ref{fig:junction_scheme-delta_n_dc}}).
Each junction waveguide has the same cross-sectional design with a 450 nm width, a 220 nm height, and a 100 nm slab thickness, suitable for the propagation of a single TE polarization mode.
The unbalancing of the MZMs is realized by a length difference \mbox{$\Delta L = 200 \: \upmu$m} between the arms leading to a passive phase shift \mbox{$\Delta\theta = 2\pi / \lambda n_g \Delta L$} with \mbox{$n_g=3.6$}, the group index of our waveguide.
The operating point of the MZM can thus be adjusted at the quadrature \mbox{($\Delta \theta = \pi/2$)} without the need of heaters by only tuning the laser wavelength around \mbox{1550 nm}.

\subsection{Measurement of the DC Kerr modulation}

\begin{figure*}
    \centering\includegraphics[width=\linewidth]{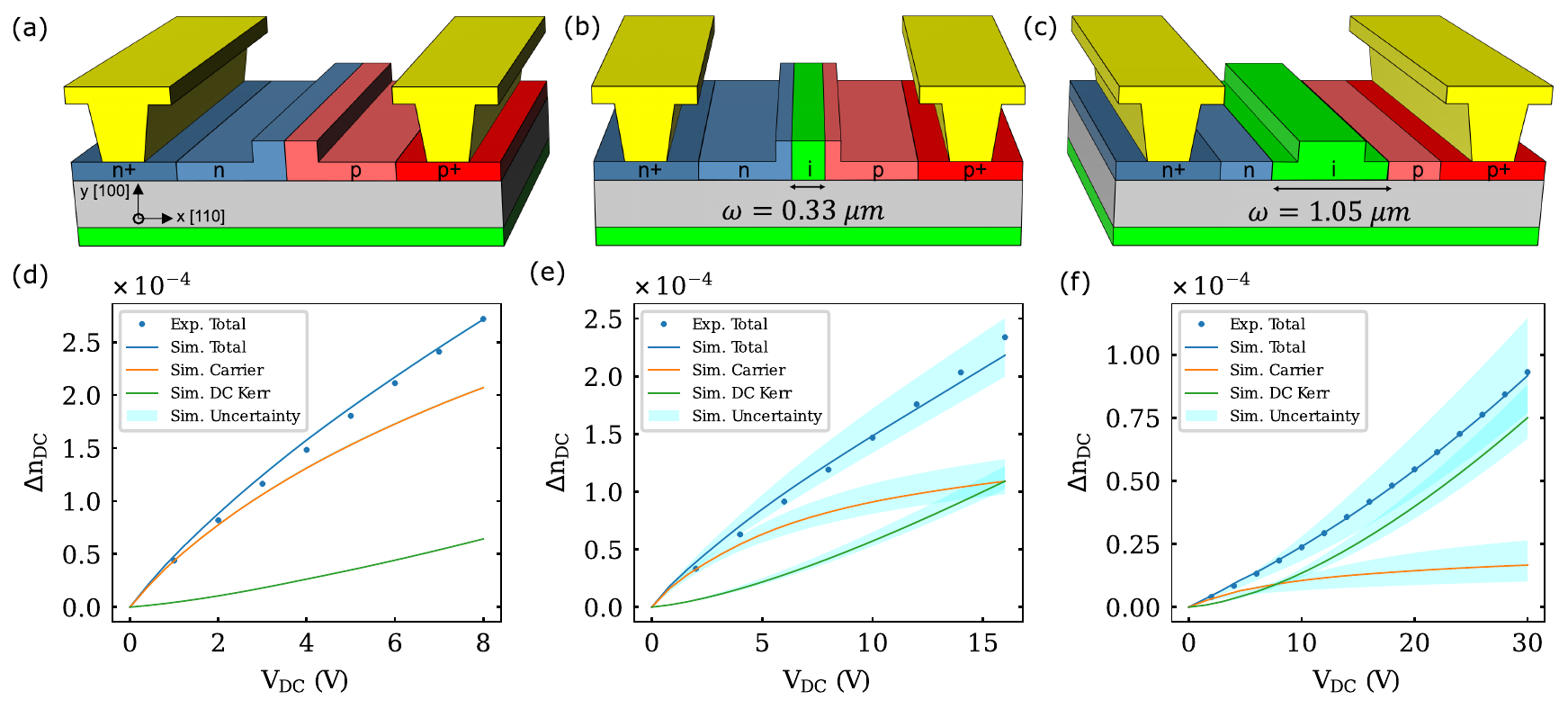}
    \caption{(a) Depiction of PN junction, (b) PIN with intrinsic region width \mbox{w = 0.33 $\upmu$m} (PIN2), and (c) PIN with \mbox{w = 1.05 $\upmu$m} (PIN3).
    \mbox{(d) Effective} refractive index changes of PN, (e) PIN2, and (f) PIN3 junctions versus the applied reverse DC bias voltage with respective MZM arm lengths of 2, 6 and 6 mm.
    Dots are the experimental measurements and lines correspond to the respective simulations of the whole modulation, of the DC Kerr and carrier modulations.}
    \label{fig:junction_scheme-delta_n_dc}
\end{figure*}

The first experiments focus on the comparison between the three junctions in MZMs under a DC bias voltage only.
The variation of the effective index of the guided mode ($\Delta n_{DC}$) as a function of the reverse DC voltage ($V_{DC}$) applied to the junction is obtained by measuring the shift of the resonance wavelength $\Delta \lambda_r$: 
\begin{equation}
    \Delta n_{DC}(V_{DC}) = \frac{\lambda_r \Delta \lambda_r(V_{DC})}{FSR(\lambda_r)  L}
    \label{eq:delta_n_dc}
\end{equation}
with $\lambda_r$ the resonance wavelength, $FSR(\lambda_r)$ the free spectral range of the MZM and $L$ the length of the electrodes all along the junctions.
See Supplement 1 section S1.
Optical and electro-optic simulations taking into account the DC Kerr and plasma dispersion effects were performed to design the three different PN/PIN waveguides.
The measured and simulated variations of the effective index of the three junctions are presented in \mbox{Fig$.$ \ref{fig:junction_scheme-delta_n_dc}}.
Total refractive index modulations are in good agreement with the simulations.
By increasing the width of the intrinsic region of the junction to 1.05 $\upmu$m, the contribution of the plasma dispersion effect is significantly reduced to become minor compared to the DC Kerr effect, while it is dominant for the PN junction waveguide.
The DC Kerr effect can thus contribute up to 82\% of the total index change in the PIN3 junction waveguide.

\subsection{Measurement of the EFI linear EO effect}
\label{EFI_section}

\begin{figure*}
    \centering\includegraphics[width=\linewidth]{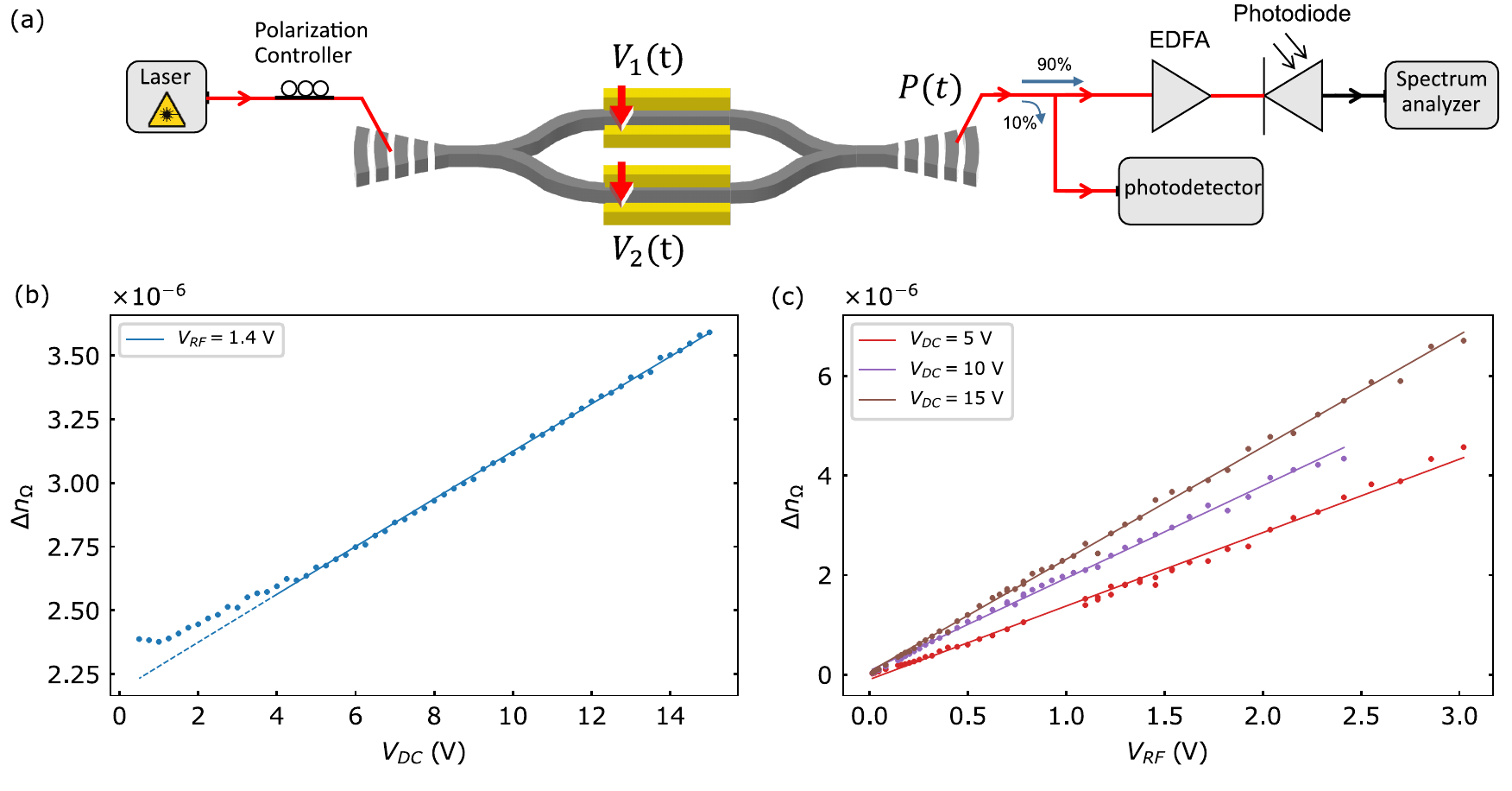}
    \caption{(a) Schematic view of the experimental setup used to measure the EOM from the MZM.
    DC voltage is applied to both arms; RF is either applied in single-drive or push-pull configuration.
    (EDFA: erbium-doped fiber amplifier).
    \mbox{(b) Effective} index variations measured in push-pull configuration versus the reverse DC bias for a fixed RF peak amplitude of 1.4V, (c) versus the RF amplitude for three reverse DC biases.}
    \label{fig:setup-m_modulation}
\end{figure*}

The study of the electric field-induced (EFI) linear EO effect in the $\Omega$ angular frequency modulation focuses on the PIN3 junction, which shows a dominant contribution of the DC Kerr effect in the effective index change (four times greater than the contribution from plasma dispersion).
A common DC bias voltage is applied to both arms of the MZM and a sinusoidal RF signal \mbox{($f = 5$ GHz)} is split with two opposite phases to be applied in push-pull configuration.
The optical wavelength is chosen to operate at the quadrature point.
A simplified schematic view of the experimental setup to characterize the EOM is provided in \mbox{Fig$.$ \ref{fig:setup-m_modulation}(a)}.
It is worthwhile to notice that the \mbox{push-pull} configuration of the MZM driving leads to assess the index variation versus voltage as an equivalent efficiency of a single path because the measured index variation is twice the index variation in each arm while the considered voltage is twice of what it is applied to each arm.
The RF analysis in push-pull configuration leads moreover to the cancellation of DC shift terms from \mbox{Eq$.$ (\ref{eq:delta_n_dc_kerr})} of the index variation in the MZM output measurements because the shift is the same in each arm.

The transfer function of the MZM as a function of the phase shift $\Delta\phi(t)$ is:
\begin{equation}
    \frac{P(t)}{P_0} = \frac{1}{2} \left \{1 + \cos[\Delta \phi(t) + \Delta \theta] \right \}
    \label{eq:TF}
\end{equation}
with $P_0$ the maximum output power of the MZM.

The EOM response at the $\Omega$ angular frequency can be approximated at the quadrature \mbox{point ($\Delta\theta = \pi/2$)} as \mbox{$P_{\Omega}(t) = $ \textonehalf $P_0 \Delta \phi(t)$} with \mbox{$\Delta \phi(t) = m_{\Omega} \cos \Omega t$}, $m_{\Omega}$ the modulation index, $m_k$ the EFI linear EO modulation index and $m_c$ the carrier modulation index:
\begin{equation}
    m_{\Omega} = m_k + m_c
    \label{eq:m}
\end{equation}
\begin{equation}
    m_k = \Gamma \frac{2 \pi}{\lambda} L_{eff1} \frac{3\chi^{(3)}}{n_{si}} F_{DC} F_{RF}
    \label{eq:mk}
\end{equation}
with the mode overlap \mbox{$\Gamma=0.87$} in the Si waveguide, the effective length \mbox{$L_{eff1} = [1-exp(-\alpha_{RF}L)] / \alpha_{RF}$} and the RF field loss \mbox{$\alpha_{RF}=4.3 \: $dB.cm$^{-1}$}.
See Supplement 1 section S2 and S3 for more details. 
Both the EFI linear EO effect and the plasma dispersion effect are expected to increase linearly with the RF amplitude.
Only the EFI linear EO effect is expected to increase with the applied reverse DC bias following the \mbox{Eq$.$ (\ref{eq:mk})}.
The dynamic carrier modulation is expected to decrease with $V_{DC}$ considering a small signal approximation on its static response.

For a 6 mm long junction, a linear behavior of the effective index change \mbox{$\Delta n_{\Omega} = m_{\Omega} \lambda / ( 2 \pi L_{eff1})$}
as a function of the applied reverse DC bias and RF amplitude is observed in \mbox{Fig$.$ \ref{fig:setup-m_modulation}(b)} and \mbox{Fig$.$ \ref{fig:setup-m_modulation}(c)}, respectively.
This is a clear signature of the EFI linear EO effect.
In \mbox{Fig$.$ \ref{fig:setup-m_modulation}(b)}, the non-zero intersection of $\Delta n_{\Omega}$ at $V_{DC}=0$ V indicates that carriers also contributed to the modulation in addition to the EFI linear EO effect at low reverse DC voltages.
The slope of the curve allows to determine the $\chi^{(3)}$ coefficient ($\chi^{(3)}=1.0 \times 10^{-19} \: $m$^2$.V$^{-2}$).
See Supplement 1 section S4 and S5 for more information.
This value is slightly underestimated (Supplement 1 section S5) due to the carriers contribution having a negative evolution with $V_{DC}$.
However, it remains relatively close to the $\chi^{(3)}$ values found in the literature.

\subsection{Measurement of the quadratic EO effect}

The quadratic EO effect at the angular frequency of 2$\Omega$ can only be observed in a single-drive configuration, as it is proportional to the square of the electric field.
We studied the transfer function at angular frequencies of $\Omega$ and 2$\Omega$ to separate the modulation behavior resulting from the distortion produced by the nonlinear transfer function of the MZM (\mbox{Eq$.$ (\ref{eq:TF})}) and the quadratic EO effect.
A bandpass RF filter centered at $\Omega$ was placed at the signal generator output insuring a very high rejection at 2$\Omega$.
We considered the PIN3 junction where distortion due to the carrier absorption modulation is negligible.

The phase shift induced by the plasma dispersion and the Kerr effects can then be written as:
\begin{equation}
	\Delta \phi (t) = m_{\Omega} \cos \Omega t + m_{2\Omega}  \cos 2\Omega t
\label{eq:phase_m2}
\end{equation}
where $m_{2\Omega}$ is the modulation index associated with the quadratic EO effect:
\begin{equation}
	m_{2\Omega} = \Gamma \frac{2 \pi}{\lambda} L_{eff2} \frac{3\chi^{(3)}}{4n_{si}} F_{RF}^2
\label{eq:m2}
\end{equation}
and \mbox{$L_{eff2} = [1-exp(-2\alpha_{RF}L)] / (2\alpha_{RF})$} is the effective length for the 2$\Omega$ component.

The $\Omega$ and 2$\Omega$ components of the MZI spectral response can be written - after inserting the phase shift $\Delta \phi (t)$ \mbox{(Eq$.$ (\ref{eq:phase_m2}))} in the MZM transfer function $P(t) / P_0$ \mbox{(Eq$.$ (\ref{eq:TF}))}, performing a Jacobi-Anger expansion and neglecting intermodulations - as follows: 
\begin{equation}
	\frac{P_\Omega (t)}{P_0} = \sin (\Delta \theta) J_1 (m_{\Omega}) \cos \Omega t
\label{eq:omega_modulation}
\end{equation}
\begin{equation}
	% \frac{P_{2\Omega} (t)}{P_0} = \left [- \cos(\Delta \theta) J_2 (m_{\Omega}) + \sin (\Delta \theta) J_0 (m_{\Omega}) J_1 (m_{2\Omega}) \right ] \cos 2\Omega t
    \begin{split}
	\frac{P_{2\Omega} (t)}{P_0} = [& - \cos(\Delta \theta) J_2 (m_{\Omega}) \\ & + \sin (\Delta \theta) J_0 (m_{\Omega}) J_1 (m_{2\Omega}) ] \cos 2\Omega t
    \end{split}
\label{eq:2omega_modulation}
\end{equation}

where $J_n (m_{\Omega})$ are the Bessel functions of the first kind.

The modulation indices $m_{\Omega}$ and $m_{2\Omega}$ are determined by fitting the DC transmission and the spectral responses using \mbox{Eq$.$ (\ref{eq:omega_modulation})} and \mbox{Eq$.$ (\ref{eq:2omega_modulation})} at a fixed reverse DC and RF voltages. See Supplement 1 section S6.

\begin{figure*}
    \centering\includegraphics[width=\linewidth]{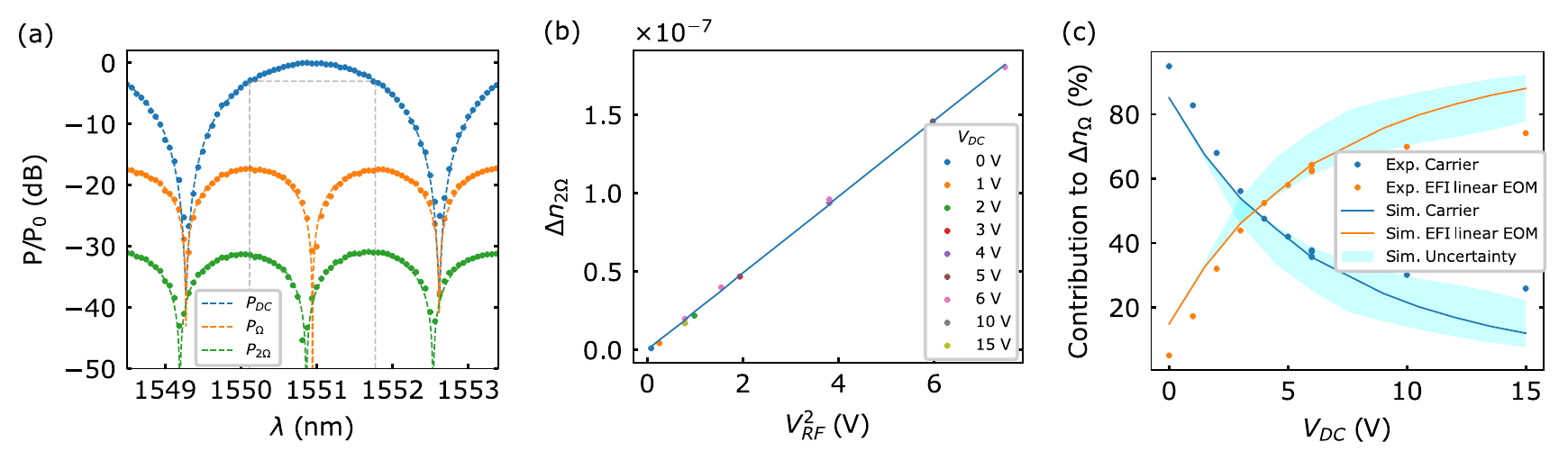}
    \caption{The dots and the lines represent respectively the measurements and the corresponding fit or simulations.
    (a) Optical MZM transfer function for three electrical spectral components excluding intrinsic losses with $P_0$ the maximum output power, $P_{DC}$ the static power, $P_{\Omega}$ the modulation power at angular frequency $\Omega$ and $P_{2\Omega}$ at frequency $2\Omega$ for the PIN3 junction by applying reverse $V_{DC}=6$ V, $V_{RF}=2.0$ V.
    (b) Amplitude of the refractive index modulation at angular frequency 2$\Omega$ versus the applied voltage $V_{RF}$ at frequency $\Omega$ for reverse DC biases from 0 to 15V.
    Whatever the value $V_{DC}$, it induces no variation of $\Delta n_{2\Omega}$.
    (c) Respective relative contribution of index variation in the $\Omega$ component from EFI linear EOM and from carrier modulation versus the applied reverse DC bias voltage.}
    \label{fig:TF_m_m2_fit_m2_contrib_mk}
\end{figure*}

The measurements performed for a 5 mm long PIN3 junction (\mbox{Fig$.$ \ref{fig:TF_m_m2_fit_m2_contrib_mk}(a)}) show that the $2\Omega$ component is induced by the quadratic EO effect and not the signal distortion (the modulation operates at quadrature).
Then, we can extract the corresponding modulation index $m_{2\Omega}$ from the response of the PIN3 junction.
We can notice that it is however not possible to extract the $m_{2\Omega}$ modulation index from the responses of the PN and PIN2 junctions because the distortion induced by carriers is too important.
See Supplement 1 section S6.
The modulation indices $m_{\Omega}$ and $m_{2\Omega}$ are accurately extracted at different reverse DC and RF bias voltages for the PIN3 junction using this method.
Experimental results are compiled in Supplementary Table S1. 

\mbox{Fig$.$ \ref{fig:TF_m_m2_fit_m2_contrib_mk}(b)} shows the linear variation of the refractive index change \mbox{$\Delta n_{2\Omega} = m_{2\Omega} \lambda / (2 \pi L_{eff2}$)} as a function of the square RF voltage (i.e. \mbox{$\Delta n_{2\Omega}$} quadratically increases with the RF voltage).
This variation is independent of the applied reverse DC voltage, as expected with a quadratic EO effect.
In addition, a linear fit of $\Delta n_{2\Omega}$ with respect to $F_{RF}^2$ is performed to extract \mbox{the $\chi^{(3)}$} coefficient \mbox{($\chi^{(3)}=1.5 \times 10^{-19} \: $m$^2$.V$^{-2}$)}. 
This value is close to the average value from the literature and is consistent with the value found in the previous section. % \ref{EFI_section}.

Moreover, the measurements of the $\Omega$ and $2\Omega$ components of the spectral response can be used to calculate the EFI linear EOM contribution to the modulation at $\Omega$ using \mbox{Eq$.$ (\ref{eq:mk})} and \mbox{Eq$.$ (\ref{eq:m2})}:
\begin{equation}
	m_k = 4\frac{F_{DC}L_{eff1}}{F_{RF}L_{eff2}} m_{2\Omega}
\label{eq:mk_from_m2}
\end{equation}

The DC electric field inside the PIN junction is estimated using \mbox{$F_{DC} = (V_{DC}+V_{bi}) / w$} with $V_{bi}$ the built-in voltage and $w$ the width of the intrinsic region \cite{Jain_2019}.
See Supplement 1 \mbox{section S4}.
The RF field is estimated from the small signal approximation \mbox{$F_{RF} \approx V_{RF} dF_{DC} / dV_{DC}$}.

The contribution of the EFI linear EOM ($m_k/m_\Omega$) and carrier modulation ($(m_\Omega-m_k)/m_\Omega$) in the $\Omega$ spectral response are reported in \mbox{Fig$.$ \ref{fig:TF_m_m2_fit_m2_contrib_mk}(c)} showing that above \mbox{$V_{DC} = 5$ V}, at a modulation frequency of 5 GHz, the EFI linear EO effect contribution to the modulation becomes greater than the carrier modulation and reaches more than a factor of 3 at 15 V.
A good agreement with simulations from \mbox{Fig$.$ \ref{fig:junction_scheme-delta_n_dc}(f)} is obtained.

\subsection{Eye diagram experiments}

\begin{figure*}
    \centering\includegraphics[width=\linewidth]{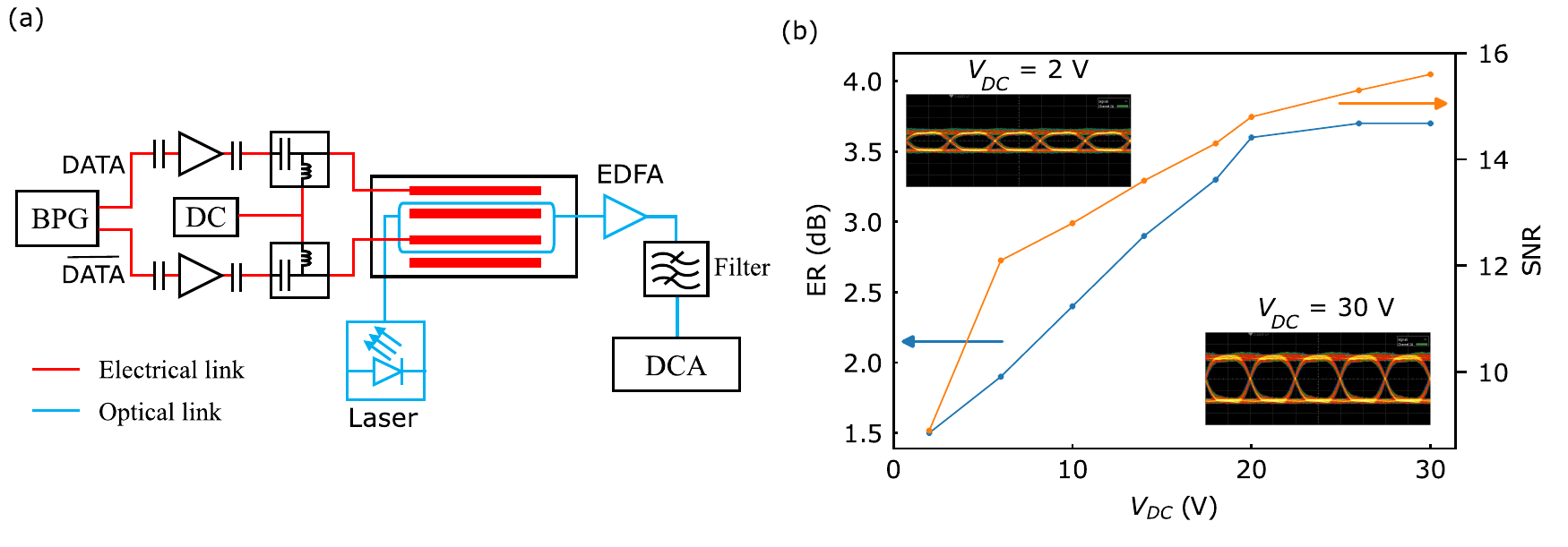}
    \caption{(a) Set-up used to acquire eye diagrams.
    (PPG: pulse pattern generator, DC: reverse DC bias, EDFA: erbium-doped fiber amplifier, DCA: digital communications analyzer).
    (b) Extinction ratio and signal to noise ratio at 10 Gbits/s by applying dual 4 V$_{pp}$ DATA/$\overline{\textnormal{DATA}}$ driving in push-pull versus the applied reverse DC bias.
    Eye diagrams for reverse DC bias of 2 and 30 V are embedded.}
    \label{fig:eye-diagrams1}
\end{figure*}

The data transmission characteristics of EO modulators based on DC Kerr effect using PIN3 diode has been analyzed.
The DATA and $\overline{\textnormal{DATA}}$ signals from a SHF bit pattern generator were amplified and transmitted to the respective arms of the MZM in push-pull configuration.
\mbox{A schematic} view of the setup is shown in \mbox{Fig$.$ \ref{fig:eye-diagrams1}(a)}.

First, optical eye diagrams were acquired at 10 Gbits/s on a digital communication analyzer (DCA) from a 6 mm long modulator with each arm driven at 4 V$_{pp}$ and at different reverse DC bias voltages.
The extinction ratio (ER) and the signal-to-noise ratio (SNR) of the modulated optical signal were computed by the DCA.
ER is greatly improved by reverse biasing $V_{DC}$ (\mbox{Fig$.$ \ref{fig:eye-diagrams1}(b)}).
Indeed, for a $V_{DC}$ varying from 2 V to 30 V, the measured ER increases from 1.5 dB to 3.7 dB, and the SNR increases from 8.9 to 15.6.
More eye diagrams as a function of $V_{DC}$ are presented in \mbox{Supplement} 1 \mbox{Fig$.$ S3}.

At higher data rate, the DC Kerr effect improves the transmission capability, reaching a maximum data rate of \mbox{40 Gbits/s} for the same \mbox{6 mm} long PIN3 modulator with each arm driven at \mbox{4 V$_{pp}$} (\mbox{Supplement 1} \mbox{Fig$.$ S4(b)}). 
\mbox{Its speed} is limited by the RF electrodes bandwidth which can be further improved by redesigning the traveling wave electrodes to achieve an expected electro-optic bandwidth of about 40 GHz for 1 cm propagation length \cite{Yang:14}.

\begin{figure}
    \centering\includegraphics[width=\linewidth]{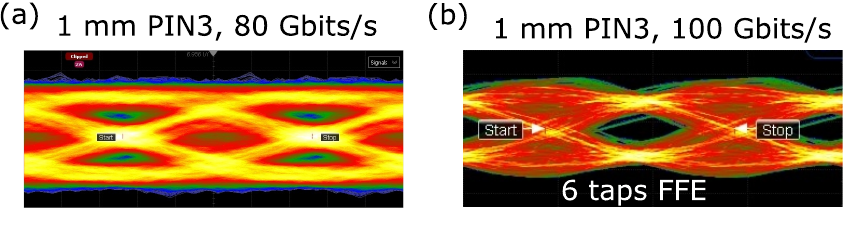}
    \caption{Optical eye diagram display from \mbox{1 mm} long PIN3 modulator by applying dual \mbox{2 V$_{pp}$} DATA/$\overline{\textnormal{DATA}}$ driving in push-pull and reverse $V_{DC}=30$ V measured at (a) 80 Gbits/s and (b) 100 Gbits/s with a numerical \mbox{6 taps} feed-forward equalization (FFE).}
    \label{fig:eye-diagrams2}
\end{figure}

Then, the bandwidth limitation of the DC Kerr effect for higher speed optical modulation has been investigated on shorter modulators with 1 mm long PIN3 modulator with each arm driven at 2 V$_{pp}$.
The obtained speed limit shows a closing of the eye diagram around 80 Gbits/s (\mbox{Fig$.$ \ref{fig:eye-diagrams2}(a)}), which is the same as the achieved speed limit of 1 mm long conventional depletion modulation under same test setup \cite{li2022112g}.
At 100 Gbits/s, the use of numerical \mbox{6 taps} feed forward equalization (FFE) has led to open the eye diagram (\mbox{Fig$.$ \ref{fig:eye-diagrams2}(b)}) showing such a DC Kerr modulator associated with the proper equalizing equipment could be promising to achieve very high speed modulation.

\section{Conclusion}

The electric field-induced Pockels effect (i.e. DC Kerr effect) has been observed in a Si PIN junction-based Mach-Zehnder modulator (MZM).
The refractive index variations as a function of both reverse DC bias voltage and RF amplitude have been measured in the dynamic regime showing a linear response with the DC bias voltage at a fixed RF amplitude.
The refractive index modulations at angular frequencies $\Omega$ and 2$\Omega$ resulting from an applied RF signal at the angular frequency $\Omega$ have been extracted to quantify the EFI linear EO effect contribution to the modulation.
We have shown that the DC Kerr effect is the main reason for the high speed modulation above 5 V DC bias voltages in comparison with plasma dispersion effect.
Furthermore, optical modulation has been demonstrated up to 100 Gbits/s for a 1 mm long Mach-Zehnder modulator.
Silicon modulators based on the electric field-induced linear EO modulation show promising characteristics for high-speed optical communications but also for applications requiring low loss and pure phase modulation.

\section{Methods}

\subsection{Sample fabrication}
The silicon MZI modulators are fabricated through silicon photonics foundry CORNERSTONE \cite{littlejohns2020cornerstone}, which provides detailed fabrication steps based on 8-inch 220 nm SOI wafers and doping information.
The passive waveguides were etched with 250 nm thickness patterned PECVD oxide hard mask.
The hard mask also protects the silicon core during the n-type implantation process.
The junction is optimized through the self-aligned doping steps in \cite{littlejohns2020cornerstone} for the studied PN and PIN junctions.

\subsection{Set-up for dynamic measurements}
A T100S-HP tunable laser is used to inject light into the device via the grating couplers.
A polarization controller is used to ensure a TE-mode injection.
A 90/10 splitter is used to separate the output power.
10\% goes into a CT400 optical components tester to measure the DC optical power and 90\% goes to a Keopsys KPS prebooster set to output a constant 3 dBm power.
The amplified modulated optical signal is collected using a Agilent 83440D photodiode and fed to an Anritsu MS2830A signal analyzer set to monitor either the $\Omega$ or 2$\Omega$ components of the spectral response.
A Keithley 2401 is used to polarized PIN junctions.
The RF signals are generated using an Anritsu MG3694C signal generator.
The signal is then coupled with the DC bias voltage using a Anritsu V251 bias-T.
For push-pull experiments, the RF signal is split in half using an Anritsu V241C power splitter and a phase delay is introduced on one arm using a Waka 02X0518-00 phase shifter.
ACP 50 GHz GSGSG RF probes are used to applied the DC and RF bias voltages to the travelling-wave electrodes.
Measurements are done at the quadrature point by tuning the laser wavelength.

\subsection{Eye diagrams experimental set-up}
MZI modulators was differentially driven with combined $V_{RF}$ and $V_{DC}$ by using two high voltage bias tees (SHF BT45R – HV100).
The high-speed signals were generated from SHF bit pattern generator and amplified to 4 $V_{pp}$ on each arm for modulations bellow 50 Gbits/s and to 2 $V_{pp}$ for higher modulations rate up to 100 Gbits/s.
NRZ signals are sent to the MZI modulators via 67 GHz GSGSG probes and terminated with DC blocks and 50m ohm resistors.
Measurements are done at the quadrature point.
Eye diagrams are displayed using the averaging function of the DCA to reduce optical noise from EDFA.

    \section*{Funding}
    EP/N013247/1, EP/T019697/1, UF150325
    
    \section*{Acknowledgment}
    The authors acknowledge CORNERSTONE team of University of Southampton for the device fabrication.
    \mbox{J. Peltier} acknowledge Victor Turpaud for fruitful discussions, and Quentin Chateiller and Bruno Garbin for the development of the Python package Autolab used in his experiments.
    This work was supported by funding from EPSRC Platform Grant (EP/N013247/1) and EPSRC Strategic Equipment Grant (EP/T019697/1).
    D. J. Thomson acknowledges funding from the Royal Society for his University Research Fellowship (UF150325).
    
    \section*{Disclosures}
    The authors declare no conflicts of interest.
    
    \section*{Data Availability}
    Data underlying the results presented in this paper are available from the corresponding authors upon reasonable request.
    
    \section*{Supplemental Document}
    See Supplement 1 for supporting content.

\end{document}

\maketitle

\section{Static measurement}

\begin{figure}
    \centering\includegraphics[width=0.5\linewidth]{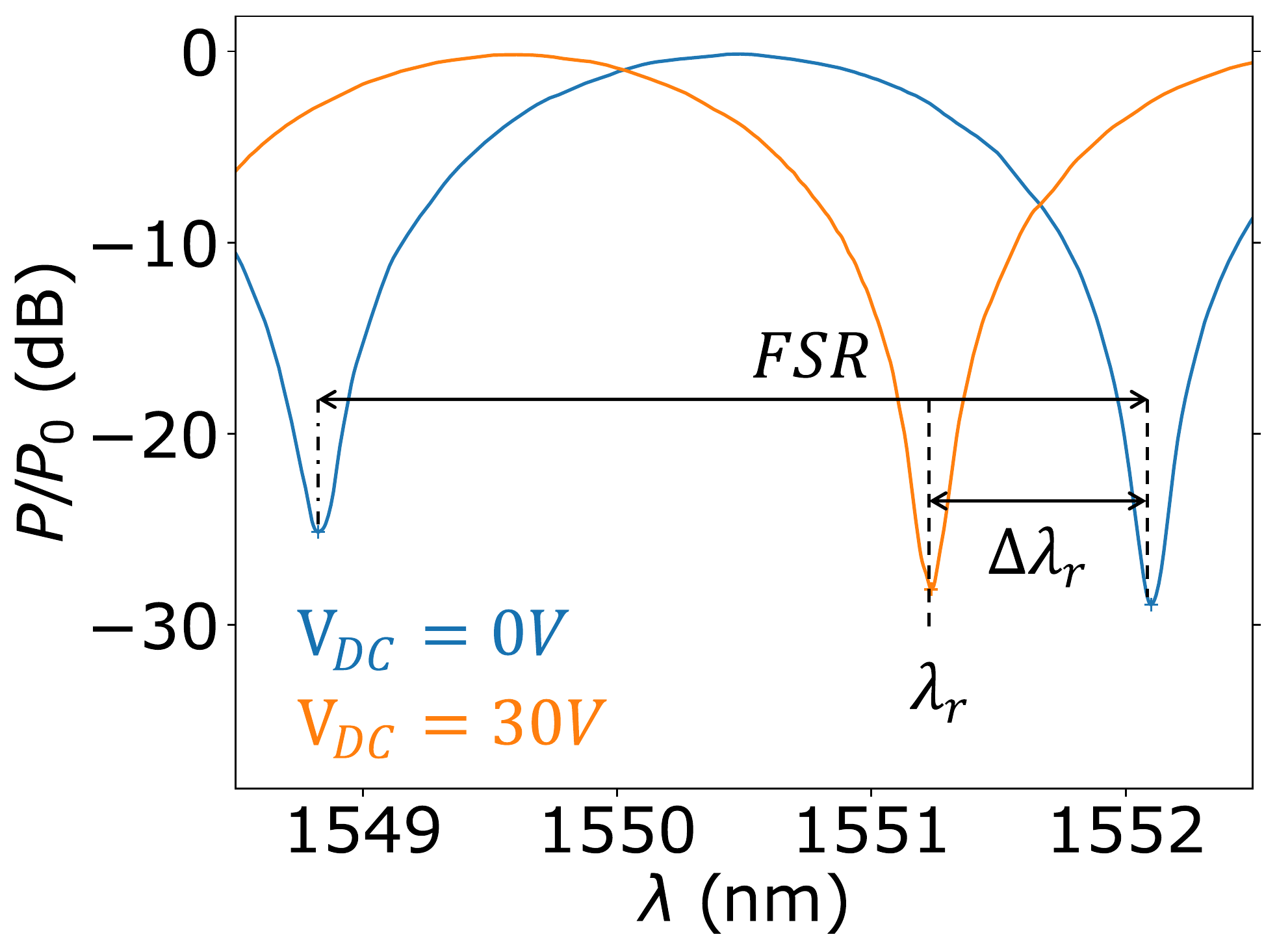}
    \caption{Optical transmission of an unbalanced MZI for an applied reverse bias of 0 and 30 V.
    \mbox{$\lambda_r$ is} the resonance wavelength, $FSR(\lambda_r)$ the free spectral range, and $\Delta \lambda_r$ the wavelength shift for bias voltage variation from 0 V to 30 V}
    \label{fig:suppl-DC_shift}
\end{figure}

Fig$.$ \ref{fig:suppl-DC_shift} shows the optical transfer function of the Mach-Zehnder interferometer (MZI) exhibiting the resonance wavelength $\lambda_r$, the free spectral range $FSR(\lambda_r)$ and the wavelength shift $\Delta \lambda_r$ \mbox{at 0 V} and \mbox{30 V} reverse DC bias.

\section{Effective index and confinement factor}

The confinement factor of the TE mode needs to be considered to correctly estimate the experimental value of the $\chi^{(3)}$ coefficient.

Effective refractive index definition:
\begin{equation}
    \Delta n_{eff} = \frac{2 n_{si}}{N} \iint_{wg}|E|^2 \Delta n dxdy
    \label{eq:n_eff}
\end{equation}
with:
\begin{equation}
    N = \frac{1}{c \epsilon_0} \iint_{\infty}(\textbf{E} \times \textbf{H}^*+\textbf{E}^*\times \textbf{H}).\hat{\textbf{z}}dxdy
    \label{eq:N}
\end{equation}

The confinement factor can be generally defined as:
\begin{equation}
    \Gamma = \frac{2 n_{si}}{N} \iint_{wg}|E|^2  dxdy
    \label{eq:Gamma}
\end{equation}
resulting in an effective refractive index:
\begin{equation}
    \Delta n_{eff} = \Gamma \frac{\iint_{wg}|E|^2 \Delta n  dxdy}{\iint_{wg}|E|^2  dxdy}
    \label{eq:n_neff_general}
\end{equation}
 
In the case of Kerr modulations, the refractive index becomes:
\begin{equation}
    \Delta n_{eff} = \Gamma  \frac{3\chi^{(3)}}{2n_{si}} \frac{\iint_{wg}|E|^2 F^2(t) dxdy}{\iint_{wg}|E|^2 dxdy}
    \label{eq:n_neff_general2}
\end{equation}

By assuming a constant F(t) field inside the waveguide, the effective refractive index modulation becomes:
\begin{equation}
    \Delta n_{eff} = \Gamma  \frac{3\chi^{(3)}}{2n_{si}} F^2(t)
    \label{eq:n_neff_general3}
\end{equation}

Therefore, the effective modulation of the $\Omega$ and $2\Omega$ components of the spectral response can be written as:
\begin{equation}
    \Delta n_{\Omega} = \Gamma  \frac{3\chi^{(3)}}{n_{si}} F_{DC}F_{RF} \cos \Omega t+ \Delta n_{carrier}
    \label{eq:n_neff_omega}
    \end{equation}
    \begin{equation}
    \Delta n_{2\Omega} = \Gamma  \frac{3\chi^{(3)}}{4n_{si}} F_{RF}^2 \cos 2\Omega t
    \label{eq:n_neff_2omega}
\end{equation}

\section{Effective length and RF losses}
\label{seq:Leff_rfloss}

RF losses from our set-up (RF filter, cables, splitter, a phase shifter and RF probes) are measured to accurately estimate the values of the applied RF amplitude $V_{RF}$ to the PIN junction. They are taken into account in $V_{RF}$ in table S1 for the corresponding output power $P_{RF}$ displayed from the generator.

RF signal loss at the position z from the line is calculated using:
\begin{equation}
    F_{RF}(z) = F_{RF}(0) \exp(-\alpha_{RF}  z)
    \label{eq:F_rf_loss}
\end{equation}

The propagation loss of the RF line \mbox{$\alpha_{RF}=4.3 \: dB.cm^{-1} = 50 \:  m^{-1}$} was extracted from RF transmissions at different RF line length.
These losses need to be taken into account in the phase shift equation to define the effective lengths.

Phase variation equation:
\begin{equation}
    \Delta \phi = \frac{2\pi}{\lambda} \int_0^L  \Delta n  dz
    \label{eq:delta_phi_delta_n}
\end{equation}

The modulation index of the electric field-induced (EFI) linear electro-optic (EO) effect at the $\Omega$ spectral component is calculated using Eq$.$ (\ref{eq:n_neff_omega}) and Eq$.$ (\ref{eq:delta_phi_delta_n}):
\begin{equation}
    m_k = \Gamma \frac{2\pi}{\lambda} \int_0^L  \frac{3\chi^{(3)}}{n_{si}}  F_{DC}F_{RF} \exp(-\alpha  z)  dz
    \label{eq:m_k_int}
\end{equation}
\begin{equation}
    m_k = \Gamma \frac{2\pi}{\lambda} L_{eff,1} \frac{3\chi^{(3)}}{n_{si}}  F_{DC}F_{RF}
    \label{eq:m_k}
\end{equation}

The effective length for this $\Omega$ component is defined as:
\begin{equation}
    L_{eff,1} = \frac{1-\exp(-\alpha_{RF} L)}{\alpha_{RF}}
    \label{eq:L_eff1}
\end{equation}

The modulation index of the quadratic EO effect at the $2\Omega$ component is calculated using \mbox{Eq$.$ (\ref{eq:n_neff_2omega})} and \mbox{Eq$.$ (\ref{eq:delta_phi_delta_n})}:
\begin{equation}
    m_{2\Omega} = \Gamma \frac{2\pi}{\lambda} \int_0^L  \frac{3\chi^{(3)}}{4n_{si}} F_{RF}^2 \exp(-2\alpha  z)  dz
    \label{eq:m_2_int}
\end{equation}
\begin{equation}
    m_{2\Omega} = \Gamma \frac{2\pi}{\lambda} L_{eff,2} \frac{3\chi^{(3)}}{4n_{si}}  F_{RF}^2
    \label{eq:m_2}
\end{equation}

The effective length for this $2\Omega$ component is defined as:
\begin{equation}
    L_{eff,2} = \frac{1-\exp(-2\alpha_{RF} L)}{2\alpha_{RF}}
    \label{eq:L_eff2}
\end{equation}

\section{Field inside the junction}
\label{seq:fields}

The DC electric field inside the PIN junction is estimated to be:
\begin{equation}
    F_{DC} = \frac{V_{DC}+V_{bi}}{w}
    \label{eq:F_DC}
\end{equation}
with the built-in voltage $V_{bi}$ being:
\begin{equation}
    V_{bi} = \frac{kT}{q} \ln{ \left(\frac{N_A N_D}{ni^2}\right)}
    \label{eq:V_bi}
\end{equation}
and the intrinsic region $w$:
\begin{equation}
    w = w_i + \sqrt{2\epsilon_0 \epsilon_{Si} e \frac{N_A+N_D}{N_A N_D}} \sqrt{V_{bi} + V_{DC}}
    \label{eq:w}
\end{equation}

For the PN3 junction, the doping level of Boron ($N_A$) in the P region and of Phosphorus ($N_D$) in the N region are \mbox{$N_A = N_D = 10^{20}$ cm$^{-3}$}.
The intrinsic region has a Boron doping level of $ni = 10^{15}$ cm$^{-3}$. $\epsilon_{Si}$=11.9 is the relative permittivity of silicon.
Resulting in $V_{bi}=0.6$ V and \mbox{$w=1050+5\sqrt{V_{DC}+0.6}$ nm.}

The RF field is estimated from the small signal approximation:
\begin{equation}
    F_{RF} \approx \frac{dF_{DC}}{dV_{DC}}V_{RF}.
    \label{eq:F_RF}
\end{equation}
Note that this approximation is particularly relevant for the PIN3 junction, even for high RF voltages, due to the small variation of the intrinsic width with the applied reverse DC bias \mbox{($dF_{DC}/dV_{DC} \approx 1/w_i$)}.

\section{Determination of the EFI linear EO effect}

In the push-pull experiment, the amplitude of the EFI linear EO modulation (at $\Omega$ component) is estimated from the slope of the DC sweep using Eq$.$ (\ref{eq:n_neff_omega}):
\begin{equation}
     \frac{d\Delta n_{\Omega}}{dF_{DC}} = \Gamma  \frac{3\chi^{(3)}}{n_{si}} F_{RF} + \Gamma  \frac{3\chi^{(3)}}{n_{si}} F_{DC} \frac{dF_{RF}}{dF_{DC}} + \frac{d\Delta n_{carrier}}{dF_{DC}}
    \label{eq:n_neff_omega_slope}
\end{equation}

For the PN3 junction, the RF field variation with the DC field is small due to the small intrinsic region width variation with the applied DC bias and can be neglected ($dF_{RF}/dF_{DC} \propto d^2F_{DC}/dV_{DC}^2 \approx 0$).
The carrier variation is however not neglectable and is expected to be negative ($d\Delta n_{carrier}/dF_{DC} < 0$), resulting in an underestimation of the $\chi^{(3)}$ coefficient using the slope of the measurement:
\begin{equation}
\chi^{(3)} \ge \frac{d\Delta n_{\Omega}} {dF_{DC}} \frac{n_{si} }{3 \Gamma F_{RF}}
\label{eq:n_neff_omega_slope_approx}
\end{equation}

\section{Fitting the spectral components}
\label{seq:fits}

The output DC optical power $P_{DC}$ of the MZI and the $\Omega$ and $2\Omega$ components of the spectral response are measured as a function of the wavelength for a fixed DC bias and RF modulation.
Their respective noise is subtracted.

The expected components are:
\begin{equation}
	\frac{P_{DC}}{P_0} = \frac{1}{2} [1 + \cos (\Delta \theta) J_0 (m_{\Omega}) J_0 (m_{2\Omega})]
\label{eq:dc_modulation}
\end{equation}
\begin{equation}
	\frac{P_\Omega (t)}{P_0} = \left [-\cos (\Delta \theta) J_1(m_{\Omega}) J_1(m_{2\Omega}) + \sin (\Delta \theta) J_0(m_{2\Omega}) J_1 (m_{\Omega}) \right ]  \cos \Omega t
\label{eq:omega_modulation}
\end{equation}
\begin{equation}
	\frac{P_{2\Omega} (t)}{P_0} = \left \{ -\cos(\Delta \theta) [J_0 (m_{2\Omega}) J_2 (m_{\Omega}) - J_2(m_{\Omega}) J_2(m_{2\Omega})] + \sin (\Delta \theta) J_0 (m_{\Omega}) J_1 (m_{2\Omega}) \right \} \cos 2\Omega t
\label{eq:2omega_modulation}
\end{equation}

In the main article, these equations are approximated assuming a small modulation index $m_{2\Omega}$ resulting in $J_0(m_{2\Omega}) \approx 1$, $J_1(m_{2\Omega})$ and $J_2(m_{2\Omega}) \approx 0$.

First, we fit the static phase variation $\Delta\theta$ of the DC curve using Eq$.$ (\ref{eq:dc_modulation}) to extract the period of the MZM assuming no dynamic modulation.
This value in then used in Eq$.$ (\ref{eq:omega_modulation}) to fit the modulation $m_{\Omega}$ index at the $\Omega$ spectral component assuming \mbox{$m_{2\Omega}$ = 0}.
Then we find the modulation $m_{2\Omega}$ at the $2\Omega$ spectral component by fitting Eq$.$ (\ref{eq:2omega_modulation}) using the previously found parameters as initial guess.

\begin{figure}
    \centering\includegraphics[width=\linewidth]{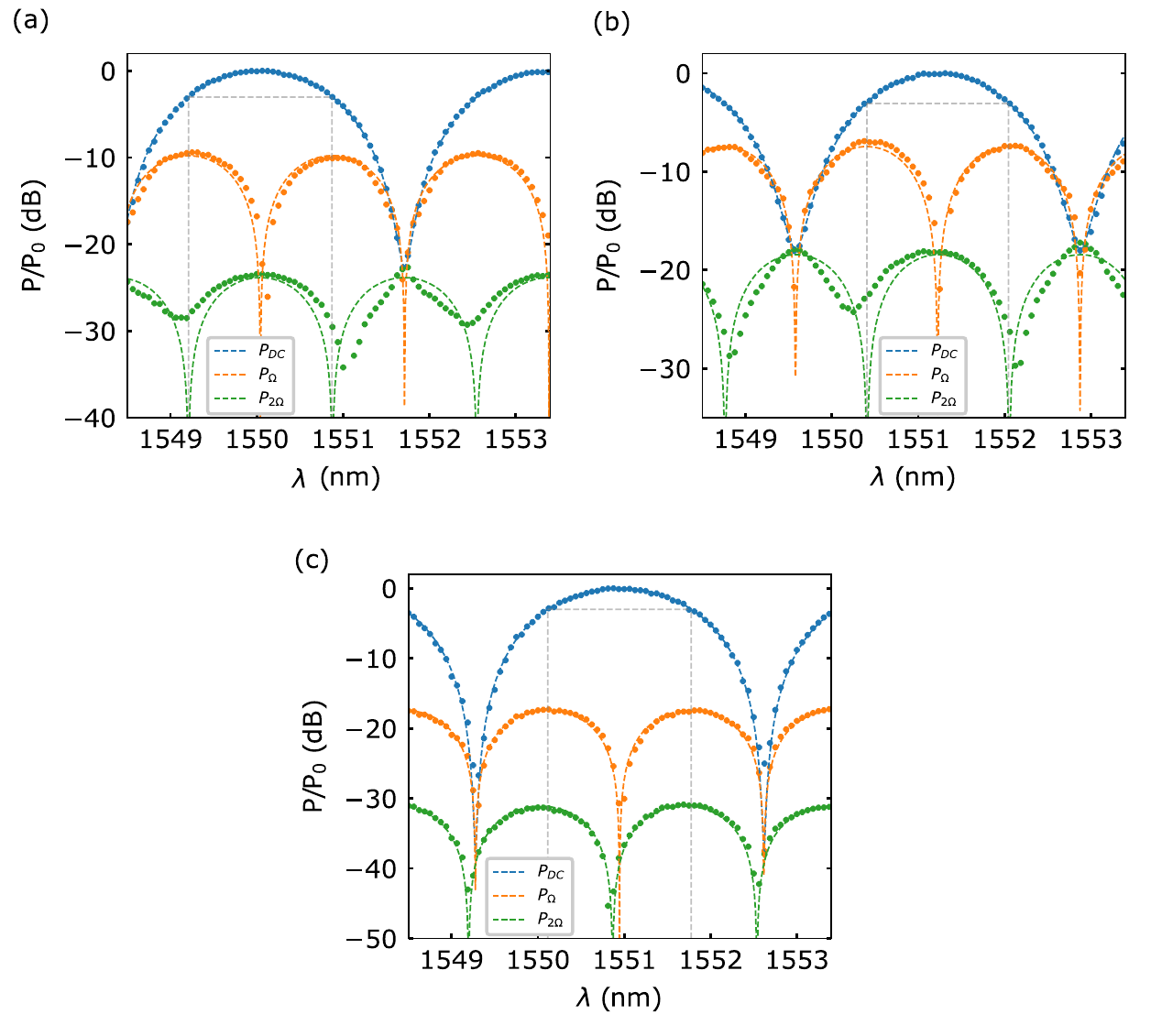}
    \caption{Measurements (dots) of $P_{DC}$ the MZM output DC optical power, $P_{\Omega}$ the modulation at angular frequency $\Omega$ and $P_{2\Omega}$ the modulation at angular frequency $2\Omega$ as a function of wavelength for 
    (a) the 2 mm long PN1 junction using $V_{DC}=2$ V, $V_{RF}=0.51$ V, 
    (b) the 5 mm long PN2 junction using $V_{DC}=4$ V, $V_{RF}=1.6$ V, 
    (c) the 5 mm long PN3 junction using $V_{DC}=6$ V, $V_{RF}=2.0$ V.
    The dashed lines represent the corresponding fit.}
    \label{fig:suppl-three_junctions_components}
\end{figure}

Fig$.$ \ref{fig:suppl-three_junctions_components} shows one of the measurements done for each of the three studied junctions.
Only the PN3 junction \mbox{(Fig$.$ \ref{fig:suppl-three_junctions_components}(c))} has a $2\Omega$ component with a modulation higher than the MZM distortion, resulting in a $2\Omega$ component aligned with the $\Omega$ component.
For the two other junctions \mbox{(Fig$.$ \ref{fig:suppl-three_junctions_components}(a) and \ref{fig:suppl-three_junctions_components}(b))}, the $2\Omega$ component comes from the distortion and the $m_{2\Omega}$ coefficient cannot be extracted.
Moreover, their spectral components do not fit as well with Eq$.$ (\ref{eq:2omega_modulation}) than for the PIN3 junction.
One possible reason could be that for those modulators, the carrier absorption introduces a chirp effect, that has not been taken into account in the analyses.

Table \ref{tab:experimental_results} regroups the modulation index $m_\Omega$ and $m_{2\Omega}$ fits for a given reverse DC bias and RF voltage.
$P_{RF}$ is the output power of the RF generator, $V_{RF}$ is the RF amplitude estimated at the beginning of the traveling-wave electrode, $F_{DC}$ and $F_{RF}$ the corresponding DC and RF field (see section \ref{seq:fields}).

\begin{table}
\centering
\caption{\bf Experimental results for the 5 mm long PIN3 junction}
\begin{tabular}{ccccccccc}
\hline
$V_{DC}$ (V) & $P_{RF}$ (dBm) & $V_{RF}$ (V) & $F_{DC}$ (V/m) & $F_{RF}$ (V/m) & $m_{\Omega}$ & $m_{2\Omega}$ & $m_k$ \\
\hline
0 & 5 & 0.29 & 5.5e+05 & 2.7e+05 & 0.0030 & 1.7e-05 & 0.00015 \\
1 & 10 & 0.51 & 1.5e+06 & 4.8e+05 & 0.0055 & 6.8e-05 & 0.00095 \\
2 & 16 & 0.99 & 2.4e+06 & 9.4e+05 & 0.013 & 0.00035 & 0.0041 \\
3 & 19 & 1.4 & 3.4e+06 & 1.3e+06 & 0.020 & 0.00075 & 0.0087 \\
4 & 22 & 2.0 & 4.3e+06 & 1.8e+06 & 0.030 & 0.0015 & 0.016 \\
5 & 24 & 2.4 & 5.3e+06 & 2.3e+06 & 0.041 & 0.0023 & 0.024 \\
6 & 15 & 0.89 & 6.2e+06 & 8.3e+05 & 0.017 & 0.00031 & 0.011 \\
6 & 18 & 1.2 & 6.2e+06 & 1.2e+06 & 0.024 & 0.00064 & 0.015 \\
6 & 22 & 2.0 & 6.2e+06 & 1.8e+06 & 0.036 & 0.0015 & 0.023 \\
6 & 25 & 2.7 & 6.2e+06 & 2.6e+06 & 0.050 & 0.0029 & 0.031 \\
10 & 15 & 0.89 & 9.9e+06 & 8.3e+05 & 0.021 & 0.00027 & 0.015 \\
15 & 15 & 0.89 & 1.5e+07 & 8.2e+05 & 0.030 & 0.00028 & 0.022 \\
\hline
\end{tabular}
  \label{tab:experimental_results}
\end{table}

\section{Eye diagrams experiments}

\begin{figure*}
    \centering\includegraphics[width=0.85\linewidth]{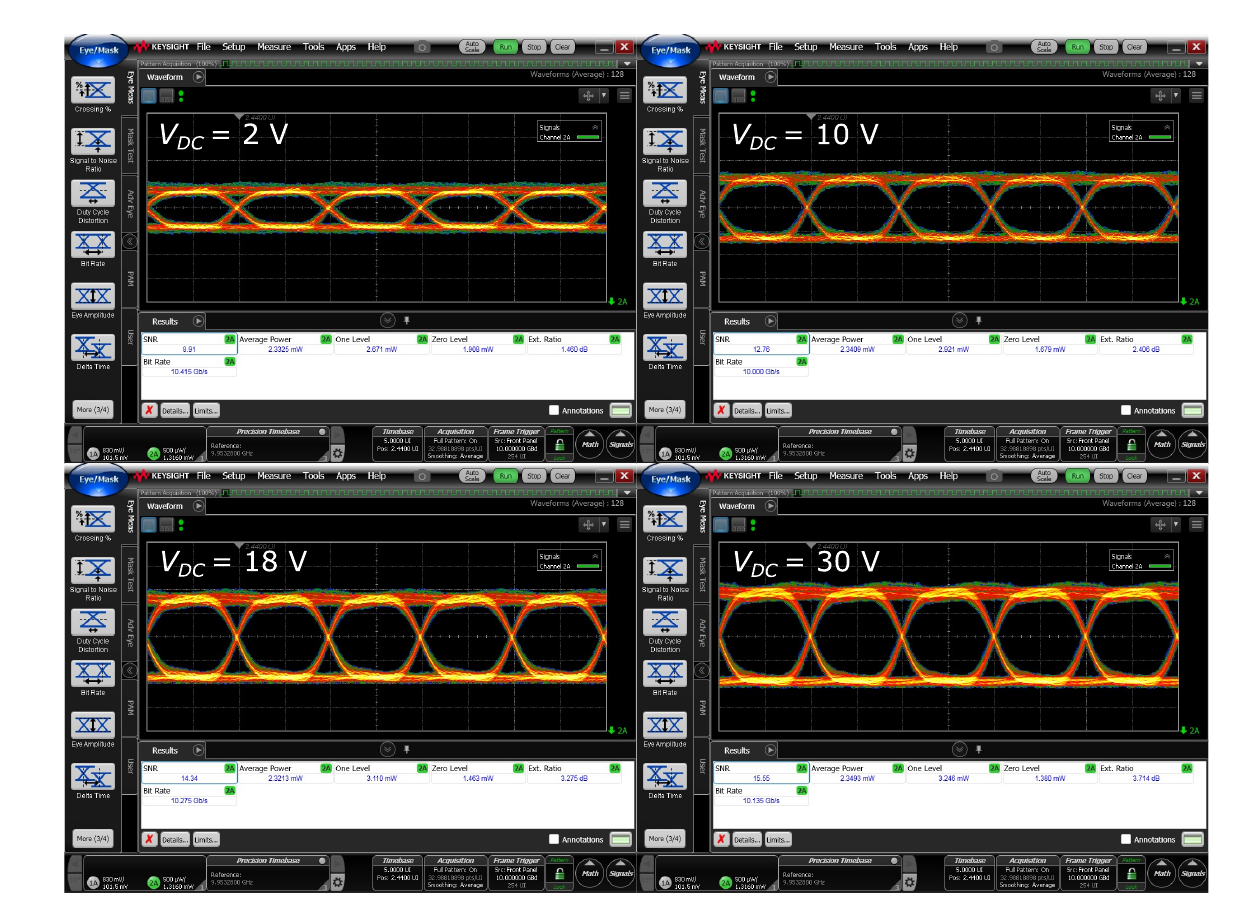}
    \caption{Eye diagram respectively measured at 2/10/18/30 V reverse DC bias at 10 Gbits/s using 4 V$_{pp}$ on each arm corresponding to ER 1.5/2.4/3.3/3.7 and SNR 8.9/12.8/14.3/15.6.}
    \label{fig:suppl-eye-diagram}
\end{figure*}

Fig$.$ \ref{fig:suppl-eye-diagram} shows some of the eye diagrams used to plot the evolution of the ER and SNR with $V_{DC}$ in the main article respectively measured at 2, 10, 18 and 30 V reverse bias for a 10 Gbits/s modulation by applying dual 4 V$_{pp}$ DATA/$\overline{\textnormal{DATA}}$ driving in push-pull.
The displays use the averaging function of the DCA to reduce optical noise from EDFA.

\begin{figure}
    \centering\includegraphics[width=0.85\linewidth]{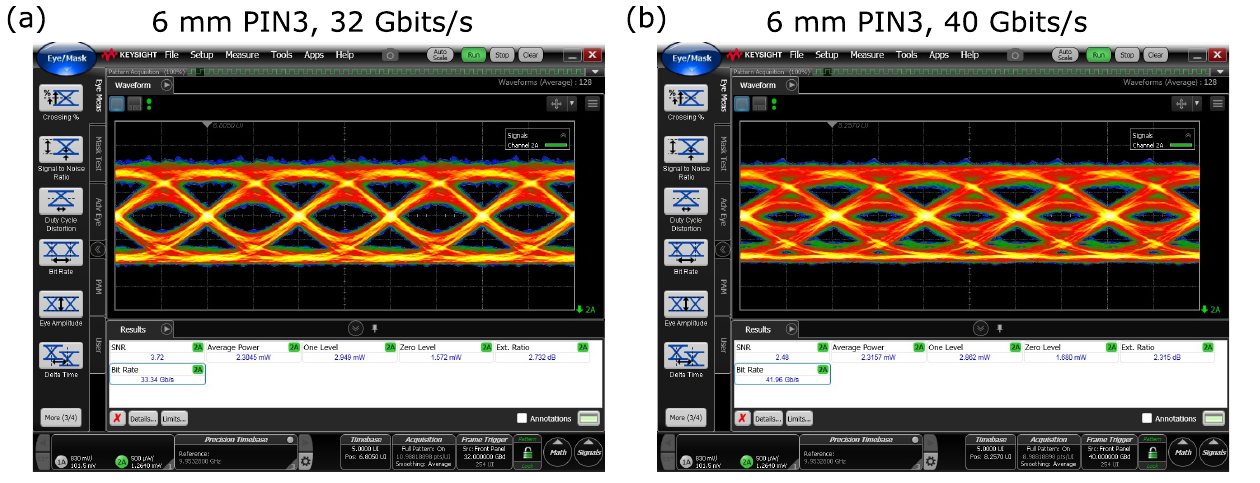}
    \caption{Optical eye diagram of 6 mm long PIN3 modulator measured at a data rate of 32 and 40 Gbits/s using 4 V$_{pp}$ on each arm and reverse $V_{DC}=30$ V, with ER 2.7 dB and 2.3 dB, respectively.}
    \label{fig:suppl-eye-diagram_32_40}
\end{figure}

Fig$.$ \ref{fig:suppl-eye-diagram_32_40}(a) shows a data rate transfer of 32 Gbits/s and Fig$.$ \ref{fig:suppl-eye-diagram_32_40}(b) shows a maximum data rate transfer of 40 Gbits/s achieved with the 6 mm long PIN3 based MZI.

